\documentclass[fleqn,usenatbib]{mnras}

\usepackage{newtxtext,newtxmath}
\usepackage[T1]{fontenc}

\DeclareRobustCommand{\VAN}[3]{#2}
\let\VANthebibliography\thebibliography
\def\thebibliography{\DeclareRobustCommand{\VAN}[3]{##3}\VANthebibliography}

\usepackage{graphicx}	% Including figure files
\usepackage{amsmath}	% Advanced maths commands
\title[Polarimetry using CMOS image sensors]{PICSARR: high-precision polarimetry using CMOS image sensors}

\author[J. Bailey et al.]{
Jeremy Bailey,$^{1,2}$\thanks{E-mail: j.bailey@unsw.edu.au (JB)}
Daniel V. Cotton,$^{3,2,4}$
Ain De Horta,$^{2}$
Lucyna Kedziora-Chudczer,$^4$
Om Shastri$^{3,5}$\\
$^{1}$School of Physics, University of New South Wales, Sydney, NSW 2052, Australia.\\
$^{2}$Western Sydney University, Locked Bag 1797, Penrith-South DC, NSW 1797, Australia.\\
$^{3}$Monterey Institute for Research in Astronomy, 200 Eighth Street, Marina CA, 93933, USA.\\
$^{4}$University of Southern Queensland, Centre for Astrophysics, Toowoomba, QLD 4350, Australia.\\
$^{5}$Los Gatos High School, 20 High School Ct, Los Gatos CA, 95030, USA.
}

\date{Accepted 2023 January 20. Received 2023 January 16; in original form 2022 October 18}

\pubyear{2023}

\begin{document}
\label{firstpage}
\pagerange{\pageref{firstpage}--\pageref{lastpage}}
\maketitle

% Abstract of the paper
\begin{abstract}
We have built and tested a compact, low-cost, but very-high-performance astronomical polarimeter based on a continuously rotating half-wave plate and a high-speed imaging detector. The polarimeter is suitable for small telescopes up to $\sim$1~m in aperture. The optical system provides very high transmission over a wide wavelength range from the atmospheric UV cutoff to $\sim$1000~nm. The high-quantum-efficiency, low-noise and high-speed of the detectors enable bright stars to be observed with high-precision as well as polarization imaging of extended sources. We have measured the performance of the instrument on 20~cm and 60~cm aperture telescopes. We show some examples of the type of science possible with this instrument. The polarimeter is particularly suited to studies of the wavelength dependence and time variability of the polarization of stars and planets.
\end{abstract}

% Select between one and six entries from the list of approved keywords.
% Don't make up new ones.

%%%%%%%%%%%%%%%%%%%%%%%%%%%%%%%%%%%%%%%%%%%%%%%%%%

%%%%%%%%%%%%%%%%% BODY OF PAPER %%%%%%%%%%%%%%%%%%

\section{Introduction}

Recent results using a new generation of high-precision polarimeters have shown that intrinsic polarization among bright stars is much more common than was previously thought \citep{cotton16a}, and have revealed a range of previously unobserved polarization mechanisms. These include the detection of polarization due to rotational distortion in hot stars \citep{cotton17,bailey20b,lewis22}, the detection of broad-band linear polarization due to magnetic fields in active dwarfs \citep{cotton17b,cotton19a} and the observation of polarization due to non-radial pulsation modes in a $\beta$~Cephei star \citep{cotton22}. In addition polarization due to photospheric reflection has been confirmed to be an important contributor to the polarization of binary systems \citep{bailey19a,cotton20}.

While many of these observations were made with 4~m class telescopes the stars are sufficiently bright that most of these phenomena could be studied with an efficient and precise instrument on a much smaller telescope. One such instrument is the Miniature High Precision Polarimetry Instrument \citep[Mini-HIPPI,][]{bailey17}, which is a miniature version of the HIPPI-2 instrument \citep{bailey20a} used on larger telescopes. Both instruments use Ferroelectric Liquid Crystal (FLC) modulators operating at 500 Hz together with Photomultiplier Tube (PMT) modules as detectors. Mini-HIPPI on a 35~cm telescope has been successfully used in a number of science projects \citep{bailey19a,cotton20,bailey20b}.
However, the quantum efficiency (QE) of the PMT detectors used is low, ranging from $\sim$40\% in the blue down to 5-10$\%$ at red wavelengths. Therefore, to make more effective use of the limited light grasp of small telescopes a different polarimeter design is required.

Here we describe PICSARR (Polarimeter using Imaging CMOS Sensor And Rotating Retarder), a novel polarimeter design made possible by recent developments in detector technology. 

\section{Design Considerations}

\subsection{Detectors}

A key problem with astronomical polarimeter design has been the difficulty in obtaining detectors with the optimal properties, a combination of high sensitivity and high speed. The CCD or Charge-Coupled Device has been the detector of choice for optical astronomy for the last 40 years or so (\citealp{howell2006} and references therein). CCDs are passive-pixel devices. The pixel array contains no active electronics (i.e. transistors). Photoelectrons are accumulated in the array during exposure. Then during the readout phase, charge is shifted through the array by applying voltages to clock electrodes, until it reaches the output. Active electronics to amplify and digitise the signals are outside the array area.
The CCD can achieve the high sensitivity and low noise needed by astronomers. However, to achieve best noise performance the CCD array must be read out slowly. 

The slow readout of CCDs makes them poorly matched to the requirements of high-precision polarimetry. It has long been recognized that rapid modulation is desirable for high-precision polarimetry as it makes the instrument insensitive to variability resulting from seeing, tracking irregularities and transparency changes. Most high-precision stellar polarimeters have used rapid modulation methods. James Kemp pioneered the use of Photoelastic modulators (PEMs) for high precision polarimetry \citep{kemp81} and these devices modulate at kHz rates. A modulation frequency of 500 Hz is used in the HIPPI class polarimeters. These high rates are difficult to match with CCD detectors. An additional issue is photon noise. To achieve high polarization precision, very large numbers of photons are needed. For example, measurement of fractional polarization to 10 ppm (part-per-million) requires detection of $\sim10^{10}$ photons. With a CCD, which has a limited well depth, very large numbers of frames need to be combined to collect enough photons, and this is not feasible with slow readout devices.

For these reasons most high-precision polarimeters have used single-pixel detectors such as avalanche photodiodes  \citep{hough06,wiktorowicz08} or PMTs \citep{bailey15, bailey17, bailey20a}. While these detectors can provide the required high speed detection and work at high photon rates, this comes at the cost of low QE (PMTs) or high thermal noise (avalanche photodiodes). In addition, both these detector types incorporate electron multiplication (i.e. produce many electrons per detected photon) and therefore suffer from ``excess noise'', the noise inherent in the multiplication process. This noise can be removed by the use of photon counting, but this is not practicable at the high photon rates needed for high-precision polarimetry. As a result these detectors always fall short of photon noise limited sensitivity.

The use of imaging detectors, and particularly double image methods, where a calcite plate or Wollaston prism is used to form images of orthogonal polarization states on the detector, can to some extent, remove the need for very rapid modulation. Good precision has been achieved with the DIPOL-2 polarimeter \citep{piirola14} that uses CCDs and modulates on timescales of $\sim$1 minute. However, the photon rate issues still limit the feasibility of observing the brightest stars at high precision with such an instrument.

\subsection{Active pixel CMOS image sensors}
\label{sec:cmos}

The deficiencies of CCDs led \citet{fossum93} to propose the alternative of an ``Active Pixel Sensor'', in which each pixel cell contains active electronics (transistors) and can be read out without the need to shuffle charge through the array. He implemented such a device using CMOS (Complementary Metal-Oxide-Semiconductor)\footnote{The standard technology now used for almost all digital and some analog electronics.} technology. This type of sensor has now become the standard for most imaging applications. Billions of these devices have been manufactured and used in many consumer and industrial applications such as smartphone cameras, surveillance cameras etc. Today they are usually referred to as CMOS image sensors.

Modern CMOS image sensors can match or outperform CCDs in almost all areas of performance. They provide lower median read noise, much higher full-frame speeds, lower power consumption, lower cost, and easier integration with other electronics. They can be integrated into a ``camera-on-a-chip'' \citep{fossum97} avoiding the need for the complex control electronics often needed for astronomical CCDs \cite[e.g.][]{leach00,waller04}. The development over the last few years of back-illuminated CMOS sensors \citep[e.g.][]{cooper19} now enables peak quantum efficiencies up to $\sim$95\% comparable with the best astronomical CCDs. 

Back-illuminated CMOS sensors are available at different price points. At the high end are Scientific CMOS (SCMOS) detectors, which have typical peak QE of $\sim$95\%, median read noise of $\sim$1 electron and are available in a variety of formats with pixel sizes up to 11 $\mu$m. However, low-cost CMOS sensors developed, in particular, for the surveillance camera market, can provide comparable performance \citep{diekmann17}. In this project we use low-cost sensors made by Sony, and based on their EXMOR-R (low-noise rolling shutter devices) and STARVIS (back-illumination) technologies. These detectors match the low read noise and high speed of the SCMOS detectors, and differ mostly in their smaller pixel sizes. The small pixels are however a good match to the small telescopes for which PICSARR is designed.

\subsection{Polarization Modulators}

The second key component of a polarimeter is the modulator. As already mentioned, modulators successfully used for high precision polarimetry include PEMs \citep{kemp81,hough06,wiktorowicz08} and FLCs \citep{bailey15,bailey17,bailey20a}. While CMOS image sensors are faster than CCDs they are not fast enough to work at the rates (>10 kHz) needed by PEMs. A polarimeter combining a CMOS detector with an FLC modulator would certainly be feasible. However, FLCs suffer from instrumental polarization effects that must be corrected by additional modulation stages (e.g. a whole instrument rotation) to achieve high precision \citep{bailey15,bailey20a}, and they have limited wavelength coverage with the modulation efficiency falling steeply away from the design wavelength. Additionally, these aspects depend on materials and construction, which have not been consistent over time, resulting in varied performance characteristics \citep{bailey20a, Cotton22b}.

For PICSARR we have therefore chosen to use a mechanically rotated half-wave plate as the modulator. A major advantage is that half-wave plates can be made in superachromatic versions that are close to half-wave over a very wide wavelength range. Since the detectors also have a wide wavelength range this results in an instrument with wide-band capability from the UV to the near-IR. Rotating wave-plates have often been used in conjunction with CCDs for polarimetry, but typically this involves stepping the plate to a new position between relatively long CCD exposures. In PICSARR we use a continuously rotating half-wave plate while recording a sequence of short exposure video frames using the high-speed CMOS imager. This enables a relatively high modulation frequency as well as the ability to observe bright stars without saturation.

The use of a continuously rotating waveplate with a Wollaston prism analyser (as in PICSARR) was described by \citet{Serkowski1974} as the ``best linear polarimeter'' in the absence of achromatic variable retarders. This modulator system was first used by \citet{lyot48} in conjunction with photocells as detectors.

A polarimeter using two SCMOS cameras and a continuously rotating waveplate has been described by \citet{shrestha20}. This instrument operates at slower rotation rates and is a more complex instrument than PICSARR. It has been used on a 2-m telescope and is designed for observations of fainter objects.

\begin{figure}
    \centering
    \includegraphics[width=\columnwidth]{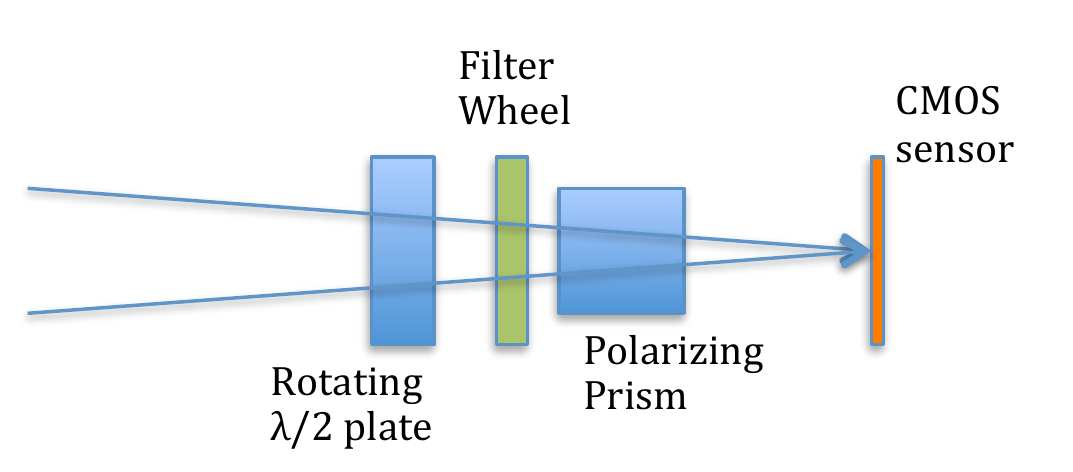}
    \caption{Optical layout of PICSARR (not to scale). All the optics are in the converging beam from the telescope. See section \ref{sec:optics} for further details.}
    \label{fig:layout}
\end{figure}

%\begin{table}
%    \caption{Filter wavelength ranges (at 50$\%$ transmission).}
%    \centering
%    \begin{tabular}{l|l}
%    \hline
%    Filter & Wavelength range (nm)  \\
%\hline     
%    u$^\prime$ & 320 -- 385 \\
%    g$^\prime$ & 401 -- 550 \\
%    r$^\prime$ & 562 -- 695 \\
%    i$^\prime$ & 695 -- 844 \\
%    z$_s^\prime$ & 826 -- 920 \\
%    Y & 950 -- 1058 \\
%    \hline
%    \end{tabular}
%    \label{tab:filters}
%\end{table}

\begin{table}
    \caption{CMOS image sensors used with PICSARR.}
    \centering
    \begin{tabular}{l|ll}
    \hline
    Camera & ASI178MM & ASI290MM \\
\hline     
    Sensor & Sony IMX178 & Sony IMX290 \\
    Format & 3096$\times$2080 & 1936$\times$1096 \\
    Pixel Size & 2.4$\mu$m & 2.9 $\mu$m \\
    Digitization & 14 / 10 bit & 12 / 10 bit \\
    Full frame rate & 60 fps & 170 fps \\
    Readout Noise (rms) & 1.4 -- 2.2 e$^-$ & 1.0 -- 3.2 e$^-$ \\
    Peak QE & 81\% & 80\% \\
    Full Well & 15 ke$^-$ & 14.6 ke$^-$ \\
    \hline
    \end{tabular}
    \label{tab:cameras}
\end{table}

\section{Instrument Description}

\subsection{Optical system}
\label{sec:optics}

The optical system of PICSARR is shown in Fig. \ref{fig:layout}. The design aims to maximize the throughput of the instrument. All of the optical components are in the converging beam from the telescope. The modulator is the first optical element in the system and is a superachromatic half-wave plate (Thorlabs SAHWP05M-700) based on the \citet{pancharatnam55} design and composed of 3 quartz and 3 magnesium fluoride plates. The waveplate has a clear aperture of 10~mm and provides a retardance within 1\% of half-wave over the wavelength range from 310 -- 1100 nm.

The polarizing prism is a magnesium fluoride Wollaston prism (Thorlabs WPM10) providing two orthogonally polarized beams with an angular separation of 1$^\circ$ 20$^\prime$. It is mounted such as to produce a double image with a separation of $\sim$ 460 $\mu$m at the detector. The use of a Wollaston prism in a converging beam results in a small focus difference between the two images, and a slight wavelength dependence in the focus. However, these effects are small compared with the typical star image sizes produced by the telescope and seeing, and the use of the best compromise focus position does not significantly limit performance.  

\begin{figure}
    \centering
    \includegraphics[width=\columnwidth]{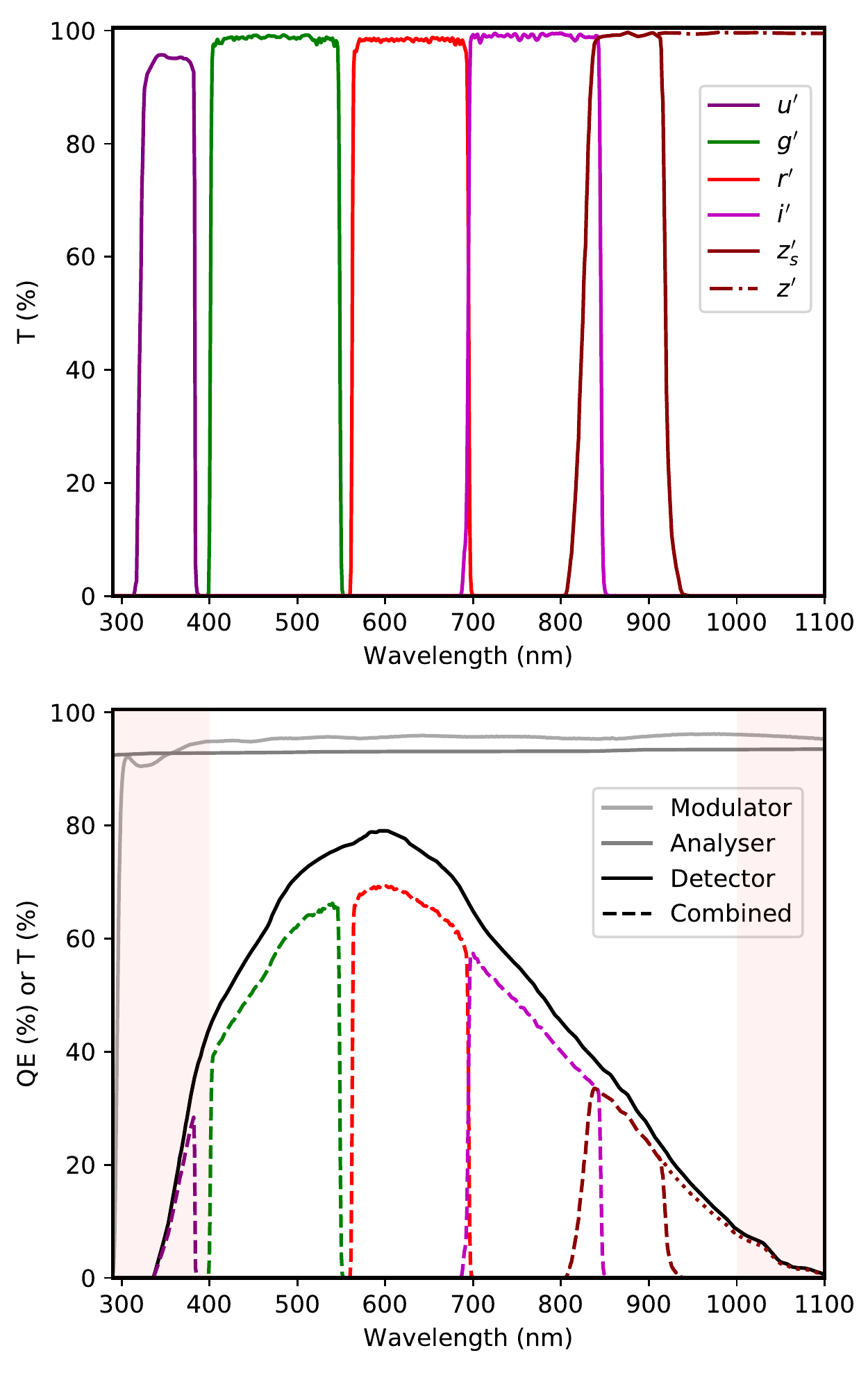}
    \caption{The upper panel shows the transmissions of the Astrodon SDSS filter set used in PICSARR. The lower panels show the contributions to the transmission from the modulator (half-wave plate), analyser (Wollaston prism) and the combined instrumental throughput including the detector QE.}
    \label{fig:tranmsission}
\end{figure}

Between the waveplate and prism there is a five position filter wheel. The filter set normally used consists of the SDSS u$^\prime$, g$^\prime$, r$^\prime$, i$^\prime$ and z$_s^\prime$ generation 2 filters from Astrodon Photometrics. The z$_s^\prime$ filter has a long wave cutoff at $\sim$920 nm, and differs from the standard z$^\prime$ filter which relies on the detector response to define the long wavelength limit. The filter transmission curves are shown in the upper panel of Fig. \ref{fig:tranmsission}.

The optical materials used for the polarization optics (quartz and magnesium fluoride), as well as the anti-reflection coating on the waveplate, result in a high throughput for these two components over the full wavelength range from 310 -- 1100 nm. The filters (g$^\prime$, r$^\prime$, i$^\prime$ and z$_s^\prime$) have transmissions of $\sim$ 98\% or better according to the manufacturer's data, and confirmed by our own laboratory measurements for the g$^\prime$, r$^\prime$ and i$^\prime$ filters \citep{bailey20a}.

\subsection{Detectors}

The detectors used with PICSARR are from the ASI range of cameras made by Suzhou ZWO Co., Ltd. The camera used for the results presented here was the ASI290MM. Some other similar format cameras such as the ASI178MM will also fit in PICSARR. These cameras use back-illuminated Sony image sensors as described in Section \ref{sec:cmos}. The specifications of these cameras are listed in Table \ref{tab:cameras}.

The cameras used in PICSARR are uncooled. While cooled versions of these cameras are also available, there is no need for cooling with the short exposure times used.  

The ASI290MM was chosen as it has the fastest readout in the ASI range, being capable of 170 frames per second (fps) with a full frame readout.  The sensor has a high speed mode with 10-bit digitization, and a slower 12-bit mode\footnote{This is referred to as 16-bit mode in the camera documentation, but only has 12 significant bits.} which is the one used by PICSARR. With the readout window reduced to 170 rows, the camera can be read out at 500~fps in 12-bit mode. 

The sensor has a peak QE of $\sim$80\% at a wavelength of 600~nm and is usable over the full range of wavelengths covered by PICSARR. The combined throughput of the instrument including the Detector QE is shown in the lower panel of Fig. \ref{fig:tranmsission}. The manufacturers data on the QE of the ASI290MM camera are only available between 400 and 1000 nm. Outside this range, data from the ASI178MM has been substituted, scaled to the ASI290MM values at the extremes. The ASI178MM uses the same protective window as the ASI290MM and this is responsible for the steep drop in transmission in the UV. The window would have to be removed or replaced to obtain the full throughput in the UV.

\begin{figure}
    \centering
    \includegraphics[width=\columnwidth]{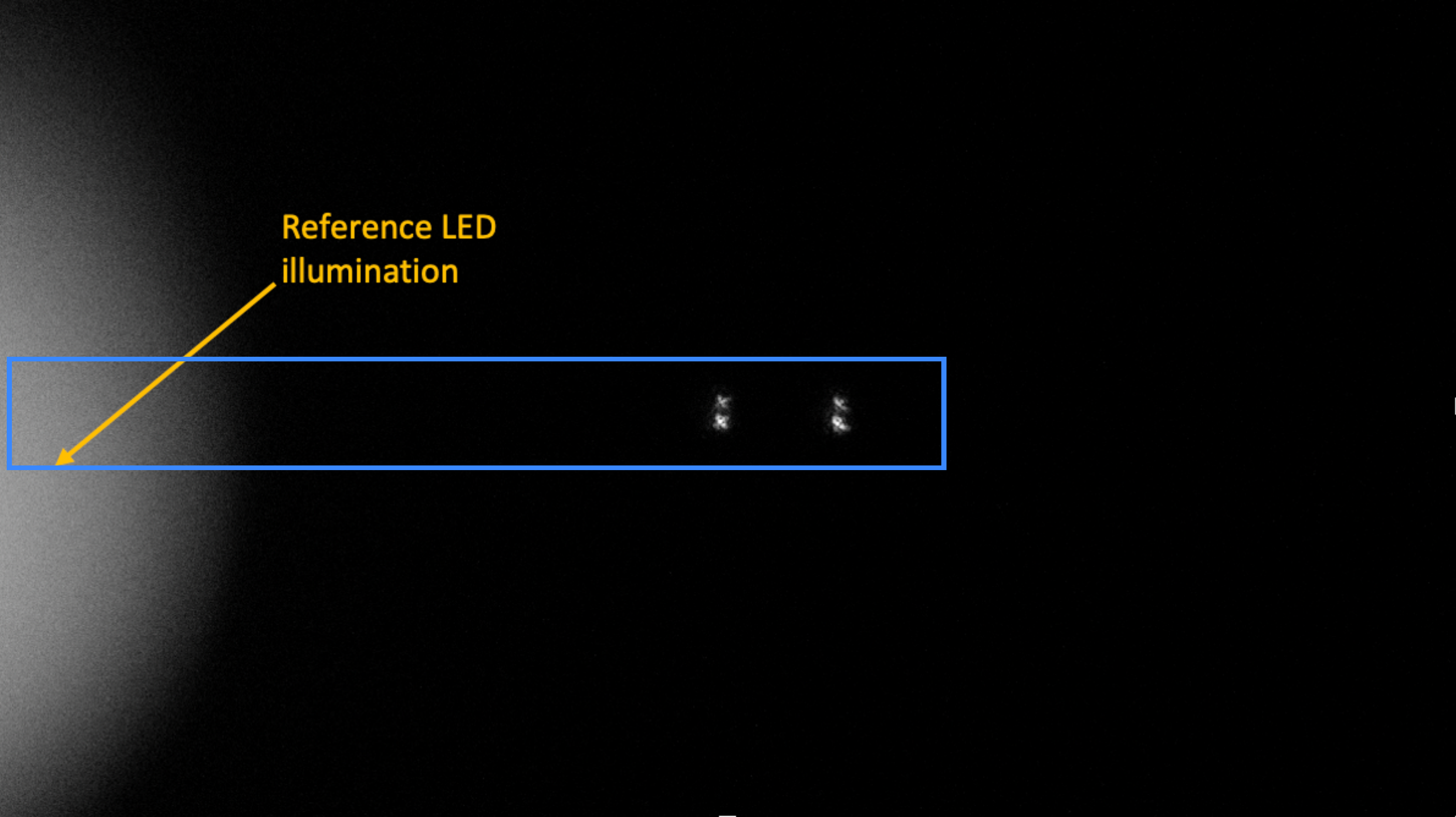}
    \caption{Raw full frame image from the ASI290MM camera showing the placement of the two star images from the Wollaston prism (here $\alpha$ Cen which is itself a double star) and the reference LED illumination. The intensity of the LED here is higher than would be needed for an actual observation. The blue box shows the 1200 $\times$ 150 readout window normally used for observation. The reference LED is at the left end, and the star data is to the right.}
    \label{fig:ref_placement}
\end{figure}

\subsection{Modulation System}

Polarization modulation is achieved by continuously rotating the half-wave plate using a Faulhaber brushless DC motor via a belt drive. The rotating waveplate results in modulation of the starlight seen at the detector with four cycles in each full rotation, and such that the amplitude and phase of the modulation provides a measure of the polarization and its position angle. The two star images produced by the Wollaston prism modulate in opposite phase, and this helps to eliminate errors in polarization due to transparency and seeing changes and non-common-path errors.

The use of a continuously rotating waveplate results in a reduced amplitude of modulation by a factor of $2\sqrt{2}/\pi$ ($\sim$0.9) compared with the case of a waveplate that is stepped to fixed positions 22.5 degrees apart. This is due to the fact that the waveplate is moving during the exposures, and the sine-wave is being integrated, rather than sampled at a fixed point. At slower speeds it would therefore by more efficient to step the waveplate between exposures but this is not practical at the rotation speeds used with PICSARR.

The waveplate rotation period is set to equal 16 frames in the video sequence recorded by the detector. This then results in each four successive frames in the video sequence sampling a full modulation cycle from which the polarization can be derived. The rotation speed is controlled by a feedback system running on a microcontroller. A Hall sensor generates a once-per-rotation pulse from the rotating waveplate, and the control system adjusts the speed of the motor to maintain the required rotation period. Rotation speeds from about 200~rpm to 2000~rpm can be selected.

The low-cost ASI cameras provide no means of electronic synchronization of the camera exposures with external events (e.g. trigger inputs). To determine the rotation phase of the waveplate we therefore use an optical reference system. The microcontroller that is used to synchronize the rotation also generates a sine-wave signal synchronized with the waveplate rotation, and uses this to drive a small light emitting diode (LED) that illuminates a region at the edge of the detector. In this way a reference signal that records the waveplate rotation phase is included in the video stream.

It is important that the reference LED signal does not contaminate the star images and impact on the polarization measurement. To ensure this, the LED illuminated spot is spaced well away from the area of the detector used for star measurement (Fig. \ref{fig:ref_placement}). Furthermore, we map the spatial distribution of scattered light from the LED and subtract this (appropriately scaled) from each frame. We are also able to select from a range of levels for the LED brightness, and adjust this to ensure that the wings of the brightness distribution are negligible at the position of the star image.

The impact of the LED reference signal is also minimized since it is at a different frequency to the polarization modulation. The LED reference signal varies at the rotation frequency, whereas the polarization modulation is at four times the rotation frequency. Since both are sine waves, the reference LED signal has no harmonics at the modulation frequency.

\begin{table}
    \caption{Standard settings for frame exposure time and rotation of the half-wave plate.}
    \centering
    \begin{tabular}{lll}
    \hline
      Frame Exp Time & Rotation Period & Rotation Rate  \\
      \hline
      12 ms & 192 ms & 312.5 rpm \\
      5 ms & 80 ms & 750 rpm \\
      2 ms & 32 ms & 1875 rpm \\
      \hline
    \end{tabular}
    \label{tab:settings}
\end{table}

\begin{figure*}
    \centering
    \includegraphics[width=17.5cm]{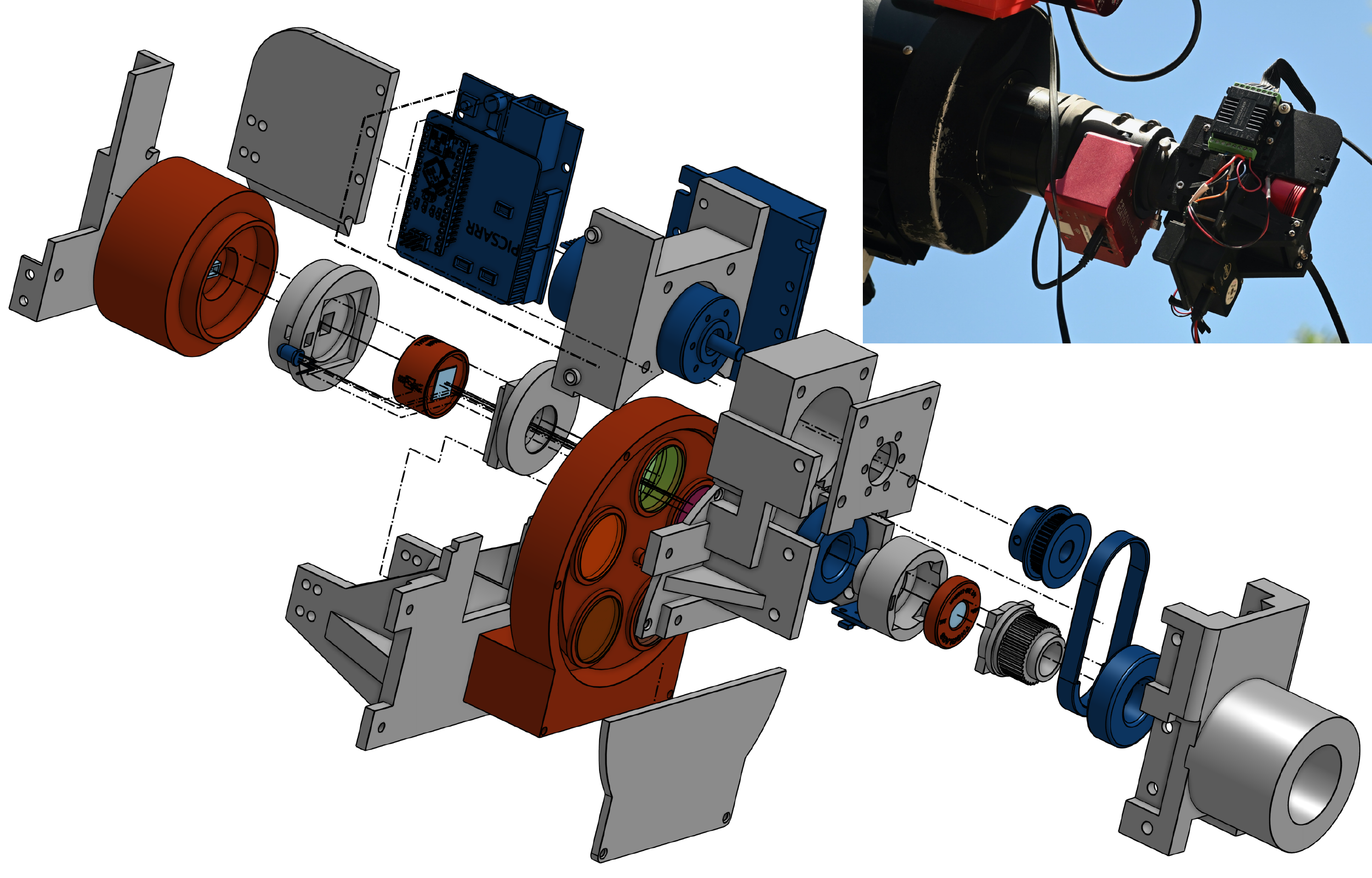}
    \caption{Exploded CAD view showing the construction of PICSARR. Grey elements are 3D printed parts. Red is used for the optical components, from left to right, the camera, Wollaston prism, filter wheel and waveplate. Blue are other parts including the motor and controller, bearings, drive belt and pulley. The inset shows PICSARR mounted on the 20cm telescope.}
    \label{fig:cad_exploded}
\end{figure*}

\subsection{Control and Data Acqusition}

Two small electronics boards are mounted on the instrument. These are the Faulhaber motor controller (type SC2804) and a custom built board that includes the microcontroller system and the synchronization and reference LED circuitry.

The instrument control and data acquisition is through software on a PC running the
Windows operating system. For the observations described here we used Intel NUC miniature PCs configured with a 2TB solid state drive (SSD). These compact, lightweight ($\sim$500~g) computers are easily mounted on even a small telescope close to the instrument. The computer is normally operated remotely over a WiFi or Ethernet connection using Microsoft Remote Desktop. The instrument requires 24~V DC power for the motor. All other power requirements are provided through the USB connections to the computer.

The camera connects to the computer using a USB 3.0 cable. Data from the camera is acquired using the SharpCap Pro\footnote{https://www.sharpcap.co.uk} image capture application running on the NUC PC. Video data is recorded to the SSD in SER ("*.ser" file) format. This format is chosen because it provides a precise time stamp for every video frame, allowing the continuity of the time sequence to be checked. 

While the system allows selection of any waveplate rotation rate within its range, we have adopted three standard settings as listed in Table \ref{tab:settings} each arranged to give 16 exposures in each waveplate rotation. The slowest of these settings (12 ms exposures and 192 ms rotation) has been the most commonly used.

A typical observation involves taking a 240 second video sequence, which, at the slowest setting, consists of about 20,000 video frames and 1250 waveplate rotations. The camera readout window is normally set to a long narrow area of 1200 by 150 pixels as shown by the blue box in Fig. \ref{fig:ref_placement}. In this case the resulting file size is 7 GB. Wider readout windows can be used, if required, to observe more extended objects.

The camera operates in a rolling shutter mode, meaning that different rows of each frame are read sequentially at slightly different times. With our standard 150 pixel wide window there is a 1.3 ms difference between the exposure mid-point time at the top and the bottom of the frame. However, all columns are read out in parallel so there is no time difference in the left-right direction. Because the double star images are spaced in the left-right direction there is no difference in their timing, and we place the star images vertically on the centre line of the window to ensure consistent phasing.

\subsection{Mechanical Construction}

The construction of PICSARR is largely by 3D printing in plastic (Acrylonitrile Butadiene Styrene or ABS). The largest part dimension is 120~mm so the parts are easily printed on a 3D printer with a small build volume. Up Plus and Zortrax M200 printers were used in the construction of the instrument. An exploded CAD drawing of the instrument is shown in Fig. \ref{fig:cad_exploded}.

The complete instrument weighs 1.2~kg and attaches to a telescope using a standard 2-inch eyepiece mount. It is therefore easily mounted on small telescopes.

\subsection{Data Reduction}

PICSARR data are reduced using software written in Python. There are two main reduction modes. For stellar objects an aperture extraction method is used to determine the polarization for data summed over a software aperture. For extended objects, the imaging reduction mode is used to obtain images in the I, Q and U Stokes parameters.

The following steps are common to both modes. Firstly each frame of the video sequence is dark subtracted and corrected for scattered light from the reference LED area.
The reference LED signal is extracted, and a sine wave fit performed to determine the waveplate rotation phase and its variation through the video sequence.

The area containing the object data is extracted, and shifts are applied to correct for image motion due to seeing and tracking errors. For stellar data the alignment is performed by locating the centre of the star images using 1D cross correlation on images collapsed in each dimension, and then moving the software aperture to centre on the star. For extended objects, frame alignment is performed using 2D cross correlation with fast Fourier transforms and then applying shifts to align the images. Only whole pixel shifts are used to avoid any need to rebin the data.

The aligned frames in each block (a block is normally one waveplate rotation consisting of 16 frames) are then used to determine the polarization and position angle for the aperture, or for each pixel in imaging mode. The position angles are then rotated based on the waveplate rotation phase determined earlier to place them on a common reference system.

In aperture mode the data for the blocks are combined to give values and statistical errors for the normalized Stokes parameters, polarization and position angle. In imaging mode the images in Stokes I, Q and U are averaged to give final images for the observation. 

For stellar observations the final stage of reduction uses software adapted from that used with the HIPPI-2 polarimeter as described by \cite{bailey20a}. This software makes use of a bandpass model for the instrument transmission as a function of wavelength, and performs final corrections to the data including calibrations for position angle zero point and telescope polarization.

The bandpass model required modifications to handle the superachromatic half-wave plate, which has a fast-axis orientation that varies with wavelength and a retardance that varies with angle of incidence. Based on the manufacturer's data and the source spectrum we make corrections to the position angle for each filter to allow for the fast-axis variations ($\sim$ a few degrees), and also make small corrections to the modulation efficiency that depend on the f-number of the telescope to allow for the angle-of-incidence dependence.

\subsection{Second Instrument}
A second PICSARR instrument has recently been built at the Monterey Institute for Research in Astronomy (MIRA). This instrument is largely the same, with the main differences designed to accommodate a larger 8 position filter wheel -- the separation between the motor and the light path is increased, requiring a longer belt.

The Astrodon filter set was unavailable, so we substituted the similar Chroma set, which has $z^\prime$ instead of $z_s^\prime$. In one of the unfilled slots we placed an opaque blank for dark calibrations. A Lenovo Nano as the computer system is the only other difference.

This instrument is currently undergoing performance evaluation on the 35-cm Celestron f/11 telescope at MIRA's Weaver Student Observatory in Marina, CA. Simultaneously it has been used in tandem with the other instrument on a number of science programs. 

\subsection{Large Telescope Use}

PICSARR is designed for small telescopes less than about 1~m in aperture. With larger telescopes, the number of pixels that would need to be combined for stellar measurements becomes very large and this increases the total readout noise. Also the spacing of the star images and the field of view for imaging become very small. However, a PICSARR-style instrument suitable for larger telescopes could be made by scaling up the optics (e.g. a larger aperture waveplate, and larger size prism) and using a sensor with larger pixels. For example, SCMOS sensors are available with 10 to 12 $\mu$m pixels. These changes would substantially increase the instrument cost.

\section{Instrument Performance}

Test observations with PICSARR have been obtained with two telescopes. The first of these is a 20~cm telescope at Pindari Observatory in suburban Sydney. This is an f/12 classical Cassegrain telescope made by Guan Sheng Optical. The second is the 60~cm f/10.5 Ritchey Chr\'{e}tien Telescope at Western Sydney University's Penrith Observatory. Due to Covid-19 restrictions, access to the Penrith Observatory has been limited, and the bulk of observations so far are with the 20~cm telescope. Table \ref{tab:telescopes} lists the pixel sizes (for the ASI290MM camera), spacing of the polarization double images, and the diffraction limited resolution for each telescope. Note that the pixel sizes provide good sampling of the diffraction limited images, and this remains true for any aperture size at similar focal ratios.

Both these telescopes use only two aluminized mirrors, and no lenses (as are included in other telescope designs such as the Schmidt Cassegrain or Corrected Dall-Kirkham). This ensures that they provide good imaging and high transmission over the wide wavelength range covered by PICSARR. They are also straight-through designs with no oblique reflections that would introduce large instrumental polarizations (e.g. as in Newtonian or Nasmyth telescopes).

A significant issue we have encountered with the use of PICSARR on the 20 cm telescope is that the use of continuously moving parts (i.e. the waveplate rotation) causes vibration in the telescope mount that appears as periodic image motion at the detector. The telescope mount (Skywatcher EQ6 Pro) is relatively lightweight, and has resonances at frequencies close to that of the waveplate rotation. We have minimized the effect by improving the alignment of the mechanical system in the instrument, and by avoiding the two faster settings (Table \ref{tab:settings}). On the 60 cm telescope which has a much more solidly built mount there is no such problem and all the settings can be used,

\subsection{Image Quality}

The combination of a small telescope, such as our 20~cm telescope, and a high-speed imaging system has a significant advantage. The telescope aperture $D$ is, under reasonable seeing conditions, only a little larger than the Fried parameter $r_{0}$ that characterizes the seeing. Under these conditions, seeing mostly shows as translational motion, and can therefore be largely eliminated by ``shift-and-add'' processing of the short exposure images \citep{smith09}.

\begin{figure}
    \centering
    \includegraphics[width=\columnwidth]{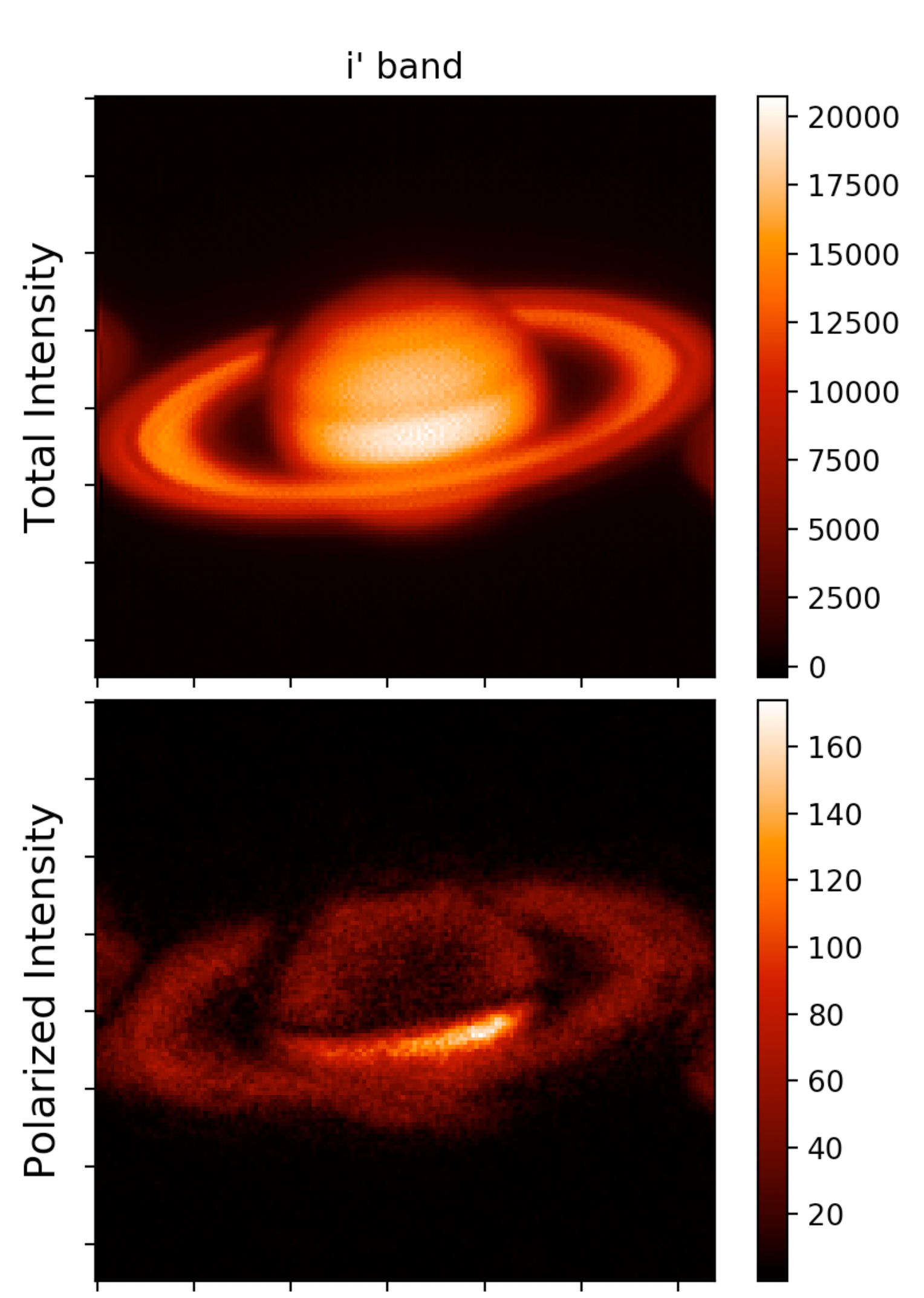}
    \caption{Total and Polarized Intensity images of Saturn  obtained with PICSARR on 2021 Sep 8th at i$^\prime$ band with the 20~cm telescope. The images demonstrate the image quality obtainable with the polarimeter. In this and similar plots the intensity units are arbitrary, but the fractional polarization can be obtained by dividing the polarized by the total intensity.}
    \label{fig:saturn}
\end{figure}

The performance of PICSARR on the 20~cm telescope is consistent with this behaviour. Under good seeing, star images in individual frames appear as diffaction limited cores with diffraction rings, and appearance is similar in processed images. An example from imaging polarimetry of Saturn is given in (Fig. \ref{fig:saturn}).

These results also highlight the fact that a small telescope can sometimes deliver better images than a larger telescope that operates in a high $D/r_{0}$ regime and is more affected by seeing. For example, compare Fig. \ref{fig:saturn} with the polarimetry of Saturn reported by \cite{schmid11}. 

\begin{table}
    \caption{Telescope Pixel Sizes (for the ASI290MM camera), Polarization Image Spacing and Resolution (Rayleigh Criterion, g$^\prime$ band) in arc seconds.}
    \centering
    \begin{tabular}{llll}
    \hline
       Telescope  & Pixel Size & Image Spacing & Resolution (g$^\prime$ band)   \\
       \hline
    20 cm f/12 &  0.245 & 39 & 0.60 \\
    60 cm f/10.5  &  0.095 & 15 & 0.20 \\
    \hline
    \end{tabular}
    \label{tab:telescopes}
\end{table}

\subsection{Instrument and Telescope Polarization}

\begin{table}
    \caption{Average low polarization star measurements showing telescope+instrument polarization zero points.}
    \centering
    \begin{tabular}{lllll}
    \hline
    Filter & \multicolumn{2}{c|}{60 cm Telescope} & \multicolumn{2}{c}{20 cm Telescope} \\
    & P (ppm) & PA (deg) & P (ppm) & PA (deg) \\
    \hline
    u$^\prime$ & 148 $\pm$ 53 & \phantom{0}74 $\pm$ 11 & 109 $\pm$ 7 & 125 $\pm$ 2\\
    g$^\prime$ & \phantom{0}30 $\pm$ 7 & \phantom{0}88 $\pm$ 6 & \phantom{0}45 $\pm$ 3 & \phantom{0}90 $\pm$ 2 \\
    r$^\prime$ & \phantom{0}32 $\pm$ 14 & 103 $\pm$ 15 & \phantom{0}40 $\pm$ 5 & \phantom{0}89 $\pm$ 3 \\
    i$^\prime$ & \phantom{0}15 $\pm$ 12 & \phantom{0}41 $\pm$ 28 & \phantom{0}45 $\pm$ 3 & 100 $\pm$ 2 \\
    z$_s^\prime$ & \phantom{0}24 $\pm$ 22 & \phantom{0}41 $\pm$ 30 & \phantom{0}46 $\pm$ 5 & 109 $\pm$ 3 \\
    \hline 
    \end{tabular}
    \label{tab:tp}
\end{table}

\begin{figure*}
    \centering
    \includegraphics[width=18cm]{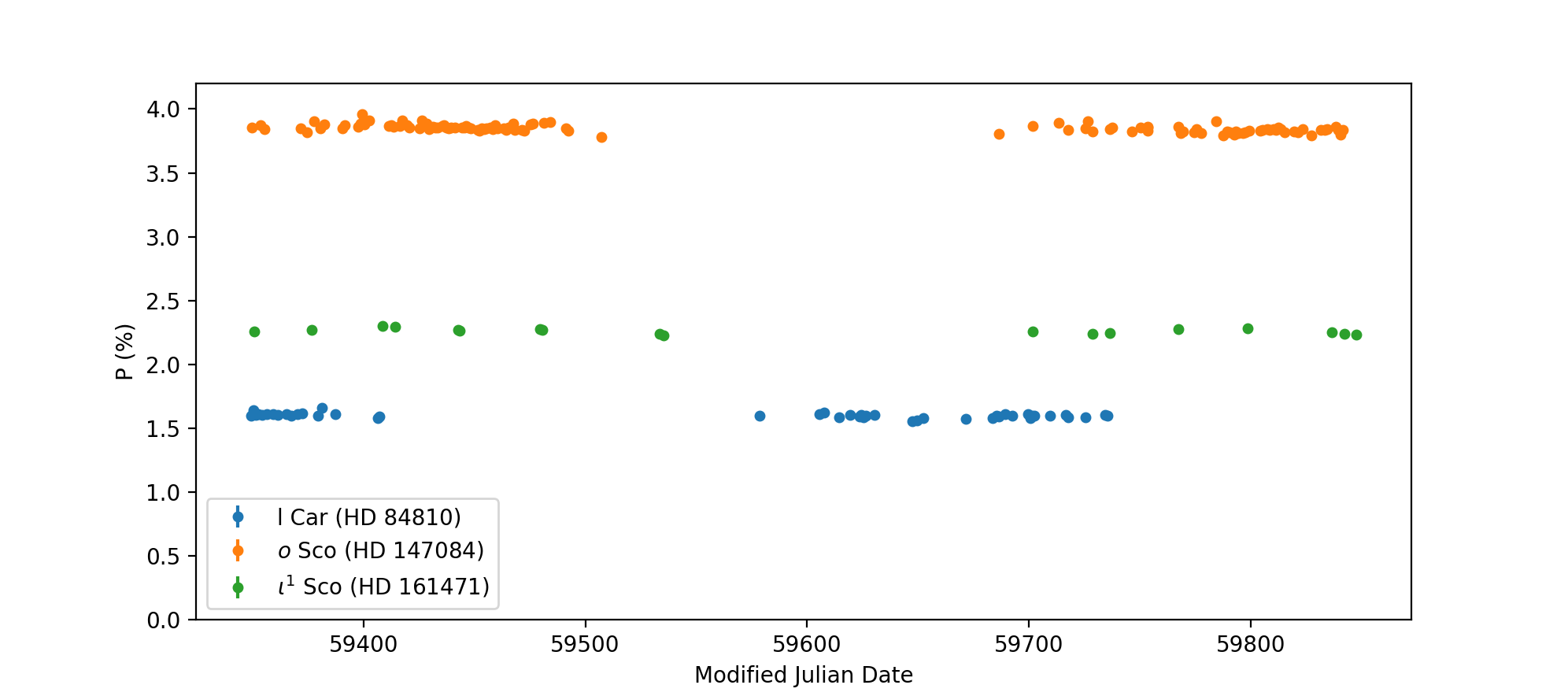}
    \caption{Polarization at g$^\prime$ for the three main high polarization standards observed with PICSARR. Most of the observations are with the 20 cm telescope. The error bars are smaller than the plotted points.}
    \label{fig:polstds}  
\end{figure*}

Measurements of nearby stars are used to determine the polarization zero point which is set by polarization introduced by the telescope optics (telescope polarization or TP) plus any instrumental polarization (IP). The low polarization standard stars used for this purpose are those listed in \cite{bailey20a}. For the 20~cm telescope, bright standard stars are needed, and the list was supplemented with the addition of Arcturus (V = $-$0.05, d = 11.3 pc) and Fomalhaut (V = 1.16, d = 7.7 pc).

The polarizations determined from observations of these stars with the 60~cm telescope in Jun 2021, and with the 20~cm telescope over May--Jun 2022 are listed in Table \ref{tab:tp}. While the PICSARR observations cannot distinguish between TP and IP, the 60~cm telescope measurement at g$^\prime$ (30 parts-per-million) agrees very well with the TP determined for this same telescope using the HIPPI-2 polarimeter \citep{bailey20a}, and thus indicates that the IP of PICSARR is small and we are measuring mostly TP.

The TP for the 20~cm telescope is then a little higher but is still less than 50 ppm, except at the extreme wavelengths of the u$^\prime$ band.

\subsection{Polarimetric Precision}

We have estimated the polarimetric precision achievable with PICSARR by looking at the night-to-night scatter of repeat measurements of low polarization stars, using the same methods as in the analysis for the HIPPI-2 instrument \citep{bailey20a}.

For the 20~cm telescope we used observations of Arcturus and Beta Leo over May--Jun 2022 from which we find a precision \citep[measured as $e_p$ as defined by][]{bailey20a}
of 17 ppm averaged over the g$^\prime$, r$^\prime$ and i$^\prime$ bands. Using observations of Sirius we found somewhat poorer precision at z$_s^\prime$ band (27 ppm) and u$^\prime$ band (40 ppm). This is excellent performance for such a small telescope.

Preliminary results from the new PICSARR instrument at MIRA on a 35 cm telescope give an $e_p$ of 11 ppm averaged over the g$^\prime$, r$^\prime$ and i$^\prime$ bands, using observations of Vega and Altair.

The limited data available with the 60~cm telescope in Jun 2021 on several low polarization standard stars gives a scatter of $\sim$10 ppm at g$^\prime$ band which is similar to the statistical precision of the individual measurements, and therefore likely an upper limit on the precision achievable. This is similar to the precision achieved on the same telescope with the HIPPI-2 polarimeter \citep{bailey20a}. The results on these three telescopes indicate improving precision with larger telescopes.

\begin{table}
    \caption{Summary of High Polarization Standard Results at g$^\prime$.}
    \centering
    \begin{tabular}{llll}
 \hline
 Star & HD 84810 & HD 161471 & HD 147084 \\   \hline
        $\lambda_{\rm eff}$ (nm) & 486.3 & 477.5 & 473.6 \\
        P (\%) & 1.596 & 2.260 & 3.850 \\
        P$_{\rm pred}$ (\%) & 1.56 & 2.20 & 3.81 \\
        $\sigma_2$ (\%) & 0.023 & 0.020 & 0.028 \\
        $\sigma_1$ (\%) & 0.005 & 0.004 & 0.008 \\
        N & 48 & 18 & 114 \\ \hline
    \end{tabular}
    \label{tab:polstds}
\end{table}

\subsection{Magnitude Range}

The useful magnitude range of PICSARR for stellar observations on the 20 cm telescope is from about magnitude 0 to 6 in the g$^\prime$ and r$^\prime$ bands. At the bright end, the limit is set by saturation of the detector during the 12 ms exposure time. This precludes observations of the few stars in the sky with negative magnitudes. The faint limit is set by the requirement that stars need to be bright enough to be located in each individual short exposure frame to allow image registration. At magnitude 6, measurements to a precision of about 100 ppm are possible.

On the 60 cm telescope it is possible to observe even the brightest stars, since the light is spread over more pixels, and since this telescope allows use of the shorter exposure settings. The faint limit is extended by another 2 magnitudes to about magnitude 8.

The faint magnitude limit might be further extended by changes to the data reduction algorithm to remove the need to realign the aperture on each frame, or modifications to the instrument that would permit slower waveplate rotation and longer frame integration times.

At other bands, limits need to be adjusted based on the star colour and detector response. At u$^\prime$ the sensitivity is much reduced since the camera includes a window that has poor transmission throughout the lower end of the band.

\subsection{High Polarization Standards}
\label{sec:polstd}

Fig. \ref{fig:polstds} shows the polarization over May 2021 to September 2022 of the three main high polarization standards observed with PICSARR at g$^\prime$. Two of these standards, HD 147084 (o Sco) and HD 84810 (l Car) are from the list given by \citet{bailey20a}. The third standard is the bright, highly polarized star HD 161471 ($\iota^1$ Sco, $V = 3.0$) for which we adopt the parameters 
$P_{\rm max} = 2.28\%$, $\lambda_{\rm max} = 0.56~\mu$m, $\theta = 2.8^\circ$ \citep*{serkowski75}. These stars are observed to calibrate the position angle of polarization. They are not used to calibrate degree of polarization, which is derived directly from the modelled behaviour of the instrument, but provide a check on the instrument stability. It can be seen that the instrument has been reasonably stable over this period. Further details on these results are given in Table \ref{tab:polstds}. The table lists the mean polarization from the observations ($P$), the predicted polarization for this filter based on the $P_{\rm max}$ and $\lambda_{\rm max}$
from \citet{bailey20a} or listed above for HD 161471. $\sigma_2$ is the standard deviation of the N observations plotted in Fig. \ref{fig:polstds}. $\sigma_1$ is the average of the internal errors of each individual observation.

Compared with the predicted values in Table \ref{tab:polstds}, the polarizations measured by PICSARR at g$^\prime$ appear slightly too high by factors of $\sim$1.01 -- 1.02. However, it should be noted that the literature values we are comparing with mostly date from the 1970s and were made with instruments with lower precision than PICSARR. They were also made in different filters, and therefore the comparison depends on correction for the wavelength dependence which introduces additional uncertainties. Further work will be needed to establish whether any additional corrections are needed to PICSARR polarization levels.

\citet{bastien88} investigated the polarization variability of polarization standard stars. They found evidence of variability in one of our standards (HD 147084) and found HD 84810 not to be variable. The methods used in this study have been criticized by \citet{clarke94}. HD 161471 is not a star that has been previously used as a polarization standard and so has not been included in past studies for variability. 

\citet{bastien88} used a filter centered on 470~nm with a width of 180~nm, which is quite similar to our g$^\prime$ filter. It is therefore interesting to compare their results for HD 147084 with ours. They find $P$ = 3.857\%, $\sigma_2$ = 0.022\%, $\sigma_1$ = 0.008\% from 41 observations, which is remarkably similar to our results of $P$ = 3.850\%, $\sigma_2$ = 0.028\%, $\sigma_1$ = 0.008\% from 114 observations\footnote{Our $\sigma_1$ and $\sigma_2$ are calculated in a different way to those of \citet{bastien88} but are roughly comparable. We could not use the Q and U measurements in the way \cite{bastien88} did as the standard observations have been used to calibrate the position angle.}. PICSARR thus achieves comparable precision despite the use of a much smaller telescope (20~cm rather than 60~cm). PICSARR could be valuable for further studies of the properties and possible variability of polarization standards, in particular, because of the potential to obtain extended series of observations.

\subsection{Sky Subtraction}

A feature of double image polarimeters (such as PICSARR and DIPOL-2) is that sky polarization is largely eliminated because the sky from the two Wollaston images overlay each other and their polarizations cancel out \citep{piirola73}. Any small residual effects resulting from asymmetry in the two channels are subsequently removed by a subtraction of the data from an annular sky region around the star aperture. This makes PICSARR very insensitive to polarized sky background. It is unaffected by moonlight. Observations can be made well into twilight. Typically stars can be observed from around 30 minutes after sunset up to 30 minutes before sunrise. Very bright objects can be observed in daylight. We have successfully measured the polarization of Venus in daylight, with Venus at less than 10 degrees away from the Sun.

Observing is also very efficient as there is no need to spend time making separate sky observations as is required with aperture polarimeters such as the HIPPI class instruments.

\begin{figure}
    \centering
    \includegraphics[width=\columnwidth]{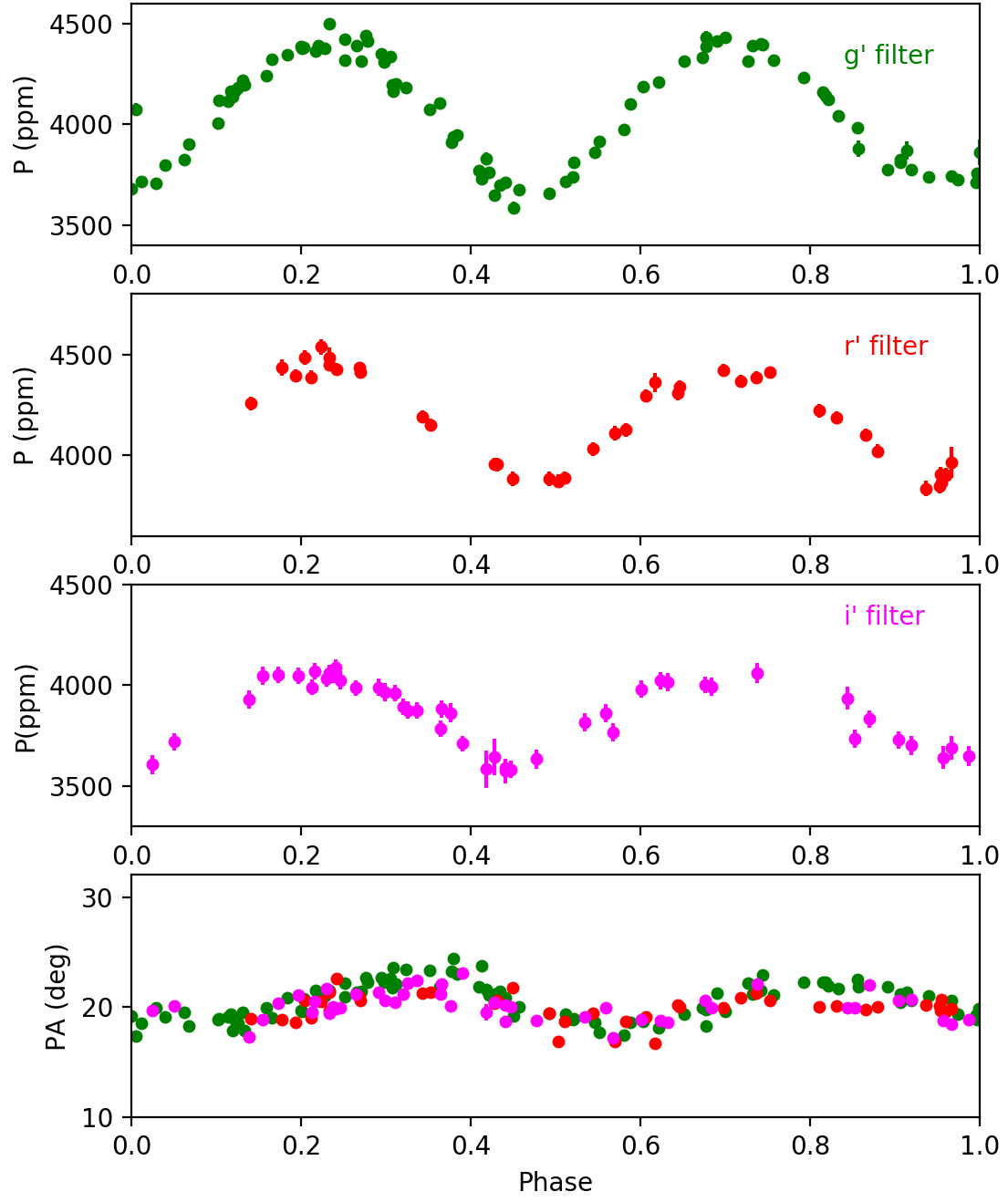}
    \caption{Polarimetry of the short period eclipsing binary $\mu^1$ Sco with PICSARR on the 20 cm telescope. The decreasing amplitude with wavelength is consistent with a photospheric reflection model for the polarization. The lower panel shows the position angle for all bands using the same colour coding as the other panels.}
    \label{fig:mu1sco}
\end{figure}

\begin{figure}
    \centering
    \includegraphics[width=\columnwidth]{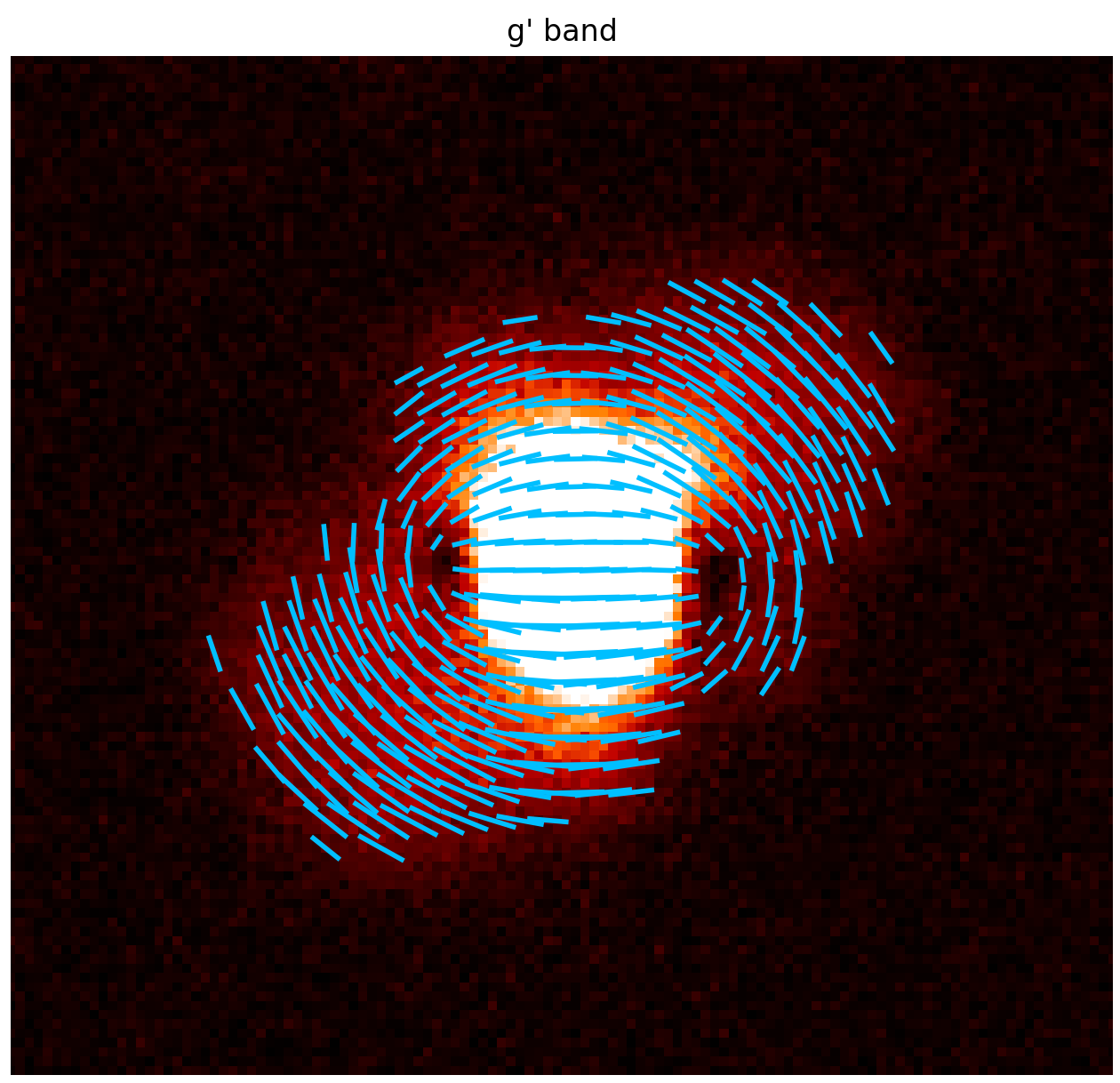}
    \caption{Polarization vectors overlaid on a polarized intensity image for the Homunculus Nebula around $\eta$ Carinae as observed with the 20 cm telescope on 2022 Feb 9th. Because the nebula is detected in polarized intensity only, the degree of polarization cannot be determined, but the vector directions are correct. The image is 27 arc seconds across. North is at top.}
    \label{fig:homunculus}
\end{figure}

\begin{figure*}
    \centering
    \includegraphics[width=16.5cm]{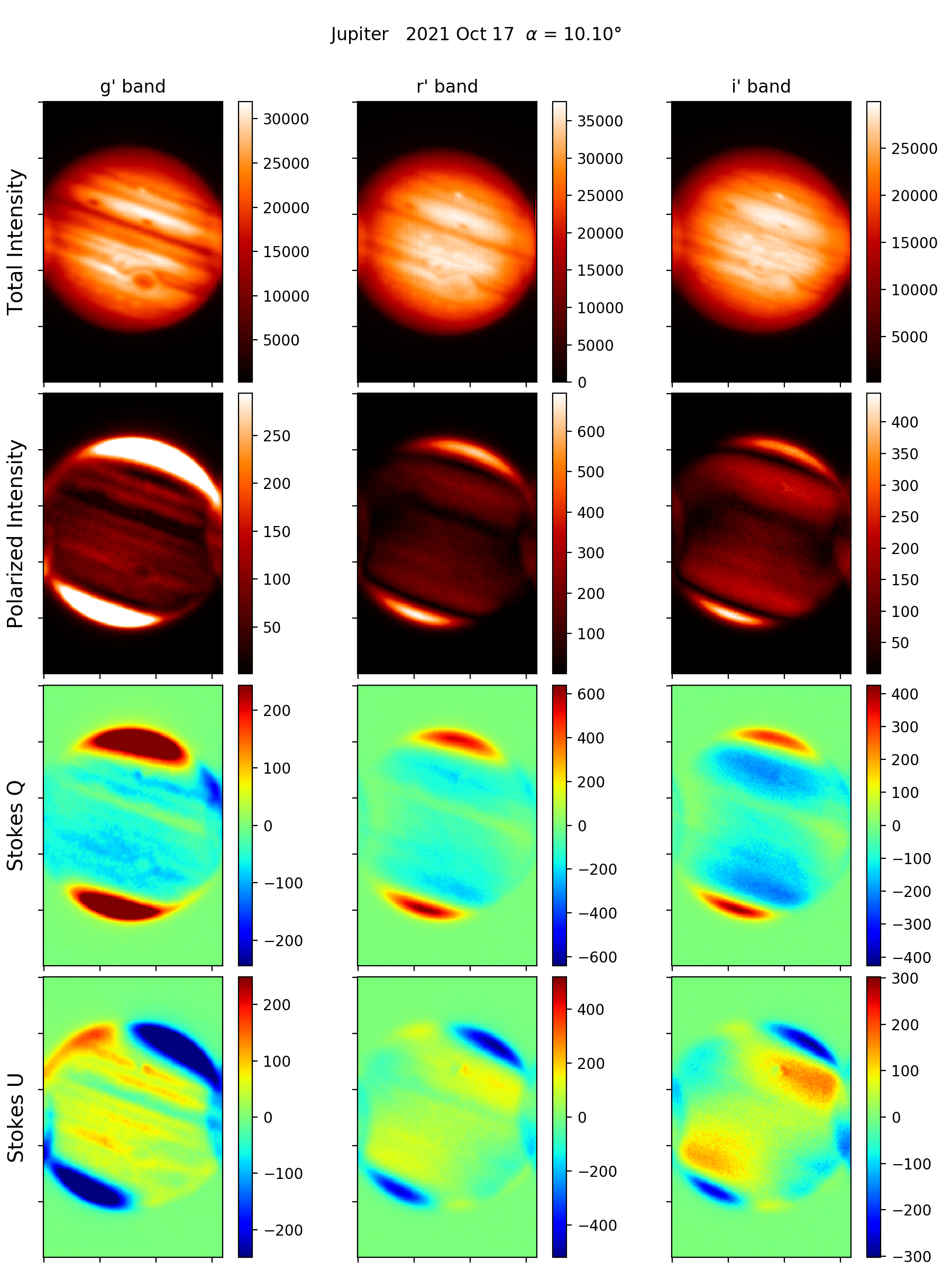}
    \caption{Imaging polarimetry of Jupiter in 3 bands obtained with PICSARR on the 20 cm telescope on 2021 Oct 17 at phase angle ($\alpha$) 10.10\degr. Note that there are some artifacts at the left and right of the images due to overlap, since the equatorial diameter of Jupiter (44 arc seconds) is greater than the double image separation from the Wollaston prism (39 arc seconds). The white spot near the top centre of the disk in the top panel is the satellite Io. It is not in transit, but overlapping the planet due to the double imaging.}
    \label{fig:jupiter}
\end{figure*}

\begin{figure}
    \centering
    \includegraphics[width=8.7cm]{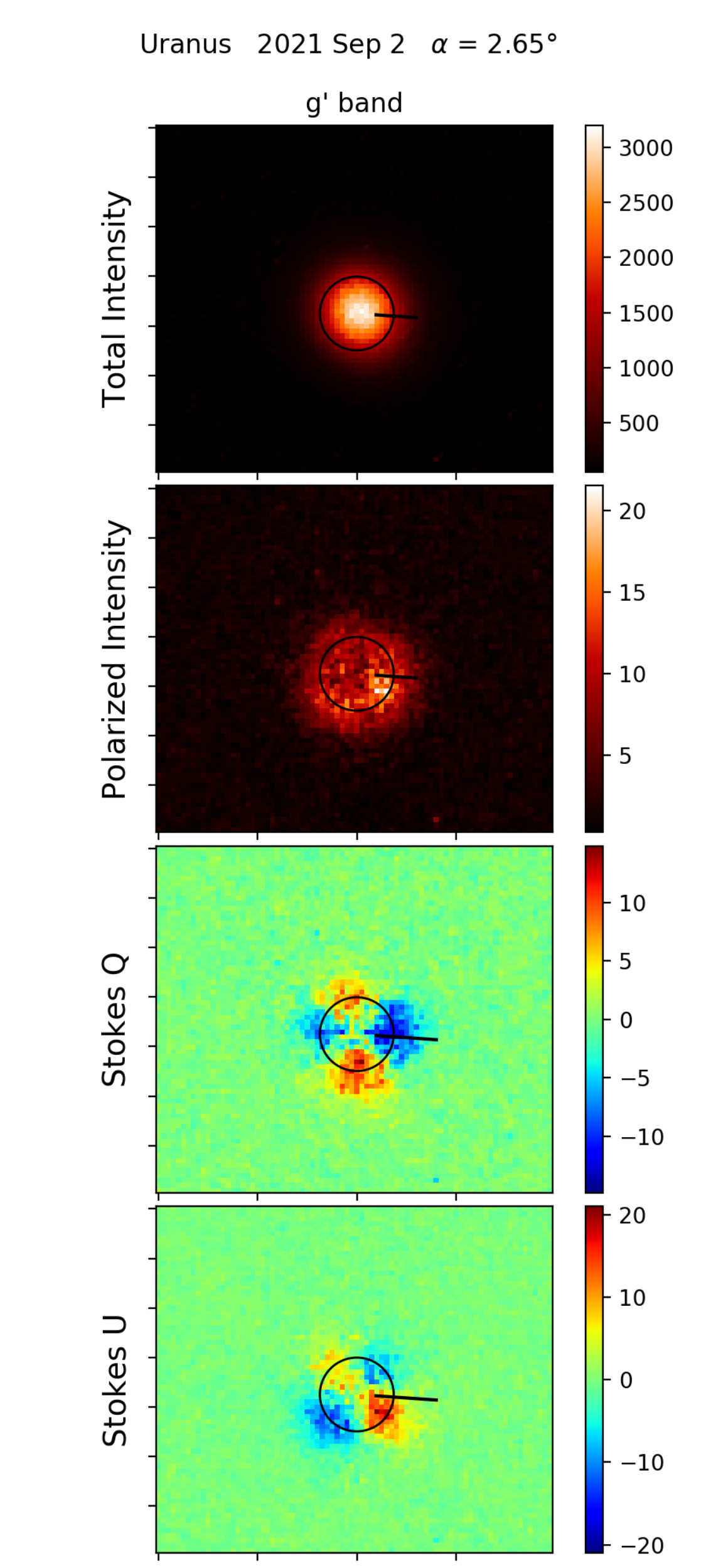}
    \caption{Imaging polarimetry of Uranus in the g$^\prime$ band obtained with PICSARR on the 20 cm telescope on 2021 Sep 2. Celestial north is at top. The circle shows the 3.65 arc second diameter of Uranus and the line shows the planetary north polar direction.}
    \label{fig:uranus}
\end{figure}

\section{Science Examples}

In this section we show some preliminary observational results that indicate the type of science that is possible with PICSARR. All these results were obtained with PICSARR on the 20~cm telescope.

\subsection{Stellar science}

Phase dependent polarization variations in close early-type binary systems have been known for some time. They have often been interpreted as due to scattering from circumstellar material resulting from the interaction between the stars \citep*{brown78}. However, there have also been suggestions that photospheric reflection between the stars can play a role in some systems \citep[e.g.][]{berdyugin99}. Recent observations \citep{bailey19a,cotton20} have provided further support for the photospheric reflection model, and showed that this model predicts that the polarization amplitude should decrease with increasing wavelength, whereas scattering from optically thin clouds would give wavelength independent polarization.

PICSARR observations of the bright short period binary system $\mu^1$ Sco are shown in Fig. \ref{fig:mu1sco} obtained with the g$^\prime$ (468 nm), r$^\prime$ (619 nm) and i$^\prime$ (754 nm) filters. The wavelengths given here are the effective wavelengths calculated by our bandpass model for this object. These results show that good quality polarization data on such objects can be obtained with PICSARR on a 20 cm telescope over a range of wavelengths. They also show decreasing amplitude of the modulation with increasing wavelength as predicted by photospheric reflection models \citep{cotton20}.

As an example of imaging polarimetry with PICSARR, Fig. \ref{fig:homunculus} shows the polarization of the Homunculus reflection nebula around $\eta$~Carinae. The reflection nebula cannot be detected in total flux in the PICSARR observations. However, it is easily detected in polarized flux. The pattern of polarization expected for a reflection nebula illuminated by the central stars is clearly shown. This image obtained with the 20 cm telescope can be compared with previous polarization maps obtained with the 3.9~m Anglo-Australian Telescope \citep{warrensmith79} and the Hubble Space Telescope \citep{schulteladbeck99,king02}. The central blob in this image is the dust-enshrouded central stars, and has been found with PICSARR to show time variable polarization ranging from 2.3\% to 3.1\%.

\subsection{Solar System science}

The wide wavelength range, and excellent image quality of PICSARR make it ideal for imaging polarimetry of planets. An example is shown in Fig. \ref{fig:jupiter} for the planet Jupiter obtained with the 20 cm telescope in three filters. Jupiter is slightly too large for the image separation given by our Wollaston prism resulting in some artifacts due to image overlap at the left and right edges.
Nevertheless the main polarization features such as the high polar polarization \citep{lyot29,gehrels69} are clearly seen. The g$^\prime$ polarization images have been scaled to show the structure over the main part of the disk associated with the markings such as the belts and zones and the Great Red Spot.

There are relatively few imaging polarimetry observations of Jupiter in the literature \citep[e.g.][]{schmid11,mclean17} and these have limited coverage in wavelength and phase angle. The ability to obtain wide wavelength coverage, high quality polarimetry with a small easily accessible telescope makes possible extensive studies of the phase angle dependence of the polarization structure, which provides information on cloud particle sizes and composition. Similar considerations apply to Venus, Mars and Saturn.

Fig. \ref{fig:uranus} shows imaging polarimetry of Uranus obtained on 2021 Sep 2 in the g$^\prime$ band. The polarization shows the ``four lobed'' structure in the Stokes parameters that can also be seen in the observations reported by \citet{schmid06} using EFOSC2 on the ESO 3.6~m telescope, but is here recorded using a 20~cm telescope. This pattern in the Stokes paramaters is the result of limb polarization in a radial direction which is caused by double scattering in the planet's atmosphere \citep{schmid06}.

\section{Conclusions}

We have built and tested a low-cost polarimeter using a rapidly rotating half-wave plate in conjunction with a high-speed CMOS image sensor. The simple optical system results in very high transmission, while the high QE and low noise of the detector provides high sensitivity. The instrument is usable over a wide wavelength range from the UV to the near-IR. It can be used to measure stellar polarization to high precision and to obtain imaging polarimetry of extended sources.

We present examples of the stellar and Solar system science possible with the instrument on a 20~cm telescope. In many cases the results are comparable with, or improve on, previous results obtained with much larger telescopes. There are very few examples of imaging polarimetry of Solar system objects in the literature, and many bright stars have not been observed polarimetrically in decades -- about a third of stars brighter than 3rd magnitude have no reported observations with a high precision instrument. The ability to do such science with small easily accessible telescopes -- even where there is extensive light pollution -- opens up the potential for much more widespread use of polarimetry. In particular it makes possible extensive studies of the time dependence of polarization in bright sources.

\section*{Acknowledgements}

We thank Prof. Miroslav Filipovic for providing access to the Penrith Observatory. DVC and OS thank the Friends of MIRA for their support. We thank an anonymous referee for suggestions that significantly improved the paper.

%%%%%%%%%%%%%%%%%%%%%%%%%%%%%%%%%%%%%%%%%%%%%%%%%%
\section*{Data Availability}

Performance data on the instrument are provided in Tables \ref{tab:tp} and \ref{tab:polstds}. Other data contained in the figures in the paper are preliminary results provided to illustrate the capabilities of the instrument. Fully reduced data will be made available in subsequent publications.

%%%%%%%%%%%%%%%%%%%% REFERENCES %%%%%%%%%%%%%%%%%%

% The best way to enter references is to use BibTeX:

\bibliographystyle{mnras}
\bibliography{picsarr} % if your bibtex file is called example.bib

% Alternatively you could enter them by hand, like this:
% This method is tedious and prone to error if you have lots of references
%\begin{thebibliography}{99}
%\bibitem[\protect\citeauthoryear{Author}{2012}]{Author2012}
%Author A.~N., 2013, Journal of Improbable Astronomy, 1, 1
%\bibitem[\protect\citeauthoryear{Others}{2013}]{Others2013}
%Others S., 2012, Journal of Interesting Stuff, 17, 198
%\end{thebibliography}

%%%%%%%%%%%%%%%%%%%%%%%%%%%%%%%%%%%%%%%%%%%%%%%%%%

%%%%%%%%%%%%%%%%% APPENDICES %%%%%%%%%%%%%%%%%%%%%

%\appendix

%\section{Some extra material}

%If you want to present additional material which would interrupt the flow of the main paper,
%it can be placed in an Appendix which appears after the list of references.

%%%%%%%%%%%%%%%%%%%%%%%%%%%%%%%%%%%%%%%%%%%%%%%%%%

% Don't change these lines
\bsp	% typesetting comment
\label{lastpage}
\end{document}